\documentclass[twocolumn]{IEEEtran}
\usepackage[T1]{fontenc}
\usepackage[latin9]{inputenc}
\usepackage{geometry}
\usepackage{color}
\usepackage{array}
\usepackage{multirow}
\usepackage{amsmath}
\usepackage{amsthm}
\usepackage{amssymb}
\usepackage{stackrel}
\usepackage{graphicx}
\usepackage[unicode=true,
 bookmarks=true,bookmarksnumbered=true,bookmarksopen=true,bookmarksopenlevel=1,
 breaklinks=false,pdfborder={0 0 0},pdfborderstyle={},backref=false,colorlinks=false]
 {hyperref}
\hypersetup{pdftitle={Your Title},
 pdfauthor={Your Name},
 pdfpagelayout=OneColumn, pdfnewwindow=true, pdfstartview=XYZ, plainpages=false}

\makeatletter

\providecommand{\tabularnewline}{\\}

\theoremstyle{definition}
\newtheorem{assumption}{Assumption}
\theoremstyle{plain}
\newtheorem{thm}{\protect\theoremname}

\usepackage[caption=false,font=footnotesize]{subfig}
\usepackage{cite}
\usepackage[caption=false,font=footnotesize]{subfig}
\IEEEoverridecommandlockouts

\@ifundefined{showcaptionsetup}{}{%
 \PassOptionsToPackage{caption=false}{subfig}}
\usepackage{subfig}
\makeatother

\providecommand{\theoremname}{Theorem}

\begin{document}
\title{
\global\long\def\rr{\mathbb{R}}%
Lyapunov-Based Kolmogorov-Arnold Network (KAN) Adaptive Control}
\author{Xuehui Shen, Wenqian Xue, Yixuan Wang, and Warren E. Dixon$^{*}$
\thanks{{*}Department of Mechanical and Aerospace Engineering, University
of Florida, USA Email: \{xuehuishen, w.xue, wang.yixuan, wdixon\}@ufl.edu. }\thanks{This research is supported in part by the Air Force Research Laboratory
(AFRL) under Grant FA8651-24-1-0018; in part by the Air Force Office
of Scientific Research (AFOSR) under Grant FA9550-22-1-0429. Any opinions,
findings, and conclusions or recommendations expressed in this material
are those of the author(s) and do not necessarily reflect the views
of the sponsoring agencies.}}
\maketitle
\begin{abstract}
Recent advancements in Lyapunov-based Deep Neural Networks (Lb-DNNs)
have demonstrated improved performance over shallow NNs and traditional
adaptive control for nonlinear systems with uncertain dynamics. Existing
Lb-DNNs rely on multi-layer perceptrons (MLPs), which lack interpretable
insights. As a first step towards embedding interpretable insights
in the control architecture, this paper develops the first Lyapunov-based
Kolmogorov-Arnold Networks (Lb-KAN) adaptive control method for uncertain
nonlinear systems. Unlike MLPs with deep-layer matrix multiplications,
KANs provide interpretable insights by direct functional decomposition.
In this framework, KANs are employed to approximate uncertain dynamics
and embedded into the control law, enabling visualizable functional
decomposition. The analytical update laws are constructed from a Lyapunov-based
analysis for real-time learning without prior data in a KAN architecture.
The analysis uses the distinct KAN approximation theorem to formally
bound the reconstruction error and its effect on the performance.
The update law is derived by incorporating the KAN\textquoteright s
learnable parameters into a Jacobian matrix, enabling stable, analytical,
real-time adaptation and ensuring asymptotic convergence of tracking
errors. Moreover, the Lb-KAN provides a foundation for interpretability
characteristics by achieving visualizable functional decomposition.
Simulation results demonstrate that the Lb-KAN controller reduces
the function approximation error by 20.2\% and 18.0\% over the baseline
Lb-LSTM and Lb-DNN methods, respectively.
\end{abstract}


\section{Introduction}

\thispagestyle{empty}Motivated by the shortcomings of Deep Neural
Networks (DNNs) where the weights are trained offline using numerical
optimization methods applied to fixed datasets, resulting in open-loop
function approximators, a series of Lyapunov-based Deep Neural Networks
(Lb-DNNs) with closed-loop online learning have been recently developed
\cite{Joshi.Chowdhary2019,Sun.Greene.ea2021,Patil.Le.ea2022,Patil.Le.ea.2022,Griffis.Patil.ea.2023,Griffis.Patil.ea23_2,Shen.Griffis.ea2025,arxivFallin.Nino.ea2025,Sweatland.Patil.ea2025,Hart.Griffis.ea2024}.
Results in \cite{Joshi.Chowdhary2019} and \cite{Sun.Greene.ea2021}
only update the output-layer weights in real-time, but the work in
\cite{Patil.Le.ea2022} leverages the compositional structure of deep
architectures to update all of the DNN weight estimates. Motivated
by the resulting performance guarantees, online learning, and no need
for the prior data, Lb-DNNs have been applied for physics-informed
neural network architectures \cite{Hart.Griffis.ea2024}, multi-agent
systems with graph neural networks \cite{arxivFallin.Nino.ea2025},
nonlinear stochastic dynamical systems \cite{Akbari.Nino.ea2024},
approximate dynamic programming \cite{Greene.Bell.ea.2023}, safe
learning with control barrier functions \cite{Sweatland.Patil.ea2025},
and further extended to a variety of learning architectures \cite{Patil.Le.ea.2022,Griffis.Patil.ea.2023,Griffis.Patil.ea23_2,Shen.Griffis.ea2025,Nino.Patil.2025,Akbari.Shen.2025}.

Like shallow NNs designed for control systems, Lb-DNNs yield analytical
update laws and performance guarantees and demonstrate significant
performance improvements; however, they employ multi-layer perceptron
(MLP) architectures, and therefore inherit the networks' internal
logic that obscures interpretability. Such MLP-based structures hinder
mechanistic interpretability (MI) \cite{Bereska2024} and \cite{Sharkey2025},
and exacerbate the curse of dimensionality. These limitations highlight
the need for Lb-adaptive control frameworks that overcome the architectural
opacity inherent in MLP-based designs.

Addressing the limitations of MLPs requires a paradigm shift from
passive post-hoc analysis of black-box models to active interpretability
by design. This perspective is known as active MI \cite{Zhang2021}
and \cite{Liu.Wang.ea2024}. Unlike passive MI \cite{Zhang2021},
which seeks to explain NNs after training, active MI emphasizes designing
interpretable architectures. The goal of active MI necessitates developing
models and training methods that are transparent from the outset.
This approach is exemplified by models derived from constructive theorems,
such as the Kolmogorov-Arnold Representation Theorem (KART) \cite{Kolmogorov1957}
and \cite{Braun2009}, which promise transparency through explicit
functional decompositions.

Inspired by KART, Kolmogorov-Arnold Networks (KANs) have been introduced
as a novel learning framework aimed at improving interpretability
and approximation accuracy over traditional MLPs \cite{Liu.Wang.ea2024}.
Unlike MLPs, which train weights and biases while fixing activation
functions, KANs incorporate learnable spline-based activation functions.
The fundamental difference lies in their architecture derived from
distinct theorems. The universal approximation theorem (UAT) \cite{Hornik.Stinchcombe.ea1989}
for MLPs guarantees the existence of NNs capable of approximating
arbitrary continuous functions but offers no constructive architecture,
whereas the KART is a constructive theorem \cite{Braun2009} which
explicitly decomposes the multivariate function into compositions
of univariate functions and sums. This architectural principle positions
KANs as a direct realization of active MI, a framework centered on
designing architectures that are inherently interpretable \cite{Liu.Wang.ea2024}.
Furthermore, unlike MLPs, which rely on increasing width and depth
to enhance performance at the cost of suffering from the curse of
dimensionality, KANs avoid this bottleneck. With a finite grid size,
KANs approximate functions with error bounds independent of dimensionality,
facilitating improved performance in high-dimensional nonlinear systems
\cite{Zhang.Li.ea2025} and \cite{So.Yung.ea2024}.

Recent studies have demonstrated the potential of KANs in a wide range
of learning tasks \cite{LiLiu2025,Lei.Deng.2025,Shukla.Toscano.ea2024,Koenig.Kim.ea2024,Zhang.Li.ea2025,Pal.Satish.ea2024,Wang.Sun.ea2024,So.Yung.ea2024}.
Among these emerging applications, a significant focus has centered
on leveraging KANs for modeling, approximating, and identifying differential
equations for deterministic continuous-time dynamical systems. Results
in \cite{Koenig.Kim.ea2024} have employed KANs as the backbone for
neural ordinary differential equations (KAN-ODEs) to efficiently learn
dynamical systems. Furthermore, KANs are increasingly serving as the
core of physics-informed neural networks to tackle forward and inverse
partial differential equation problems \cite{Shukla.Toscano.ea2024}
and \cite{Wang.Sun.ea2024}. Results in \cite{Pal.Satish.ea2024}
developed an equation discovery framework that leverages the interpretability
of KANs to identify the equation structures of nonlinear dynamical
systems. The first application of KANs to optimal control was introduced
in \cite{Aghaei2024}, where KANs were combined with physics-informed
learning to address continuous-time optimal control problems. However,
the aforementioned results use large offline datasets combined with
a numerical optimization routine to train the NN weights, lacking
real-time weight updates or provable error convergence.

The novel contribution of this work is that we develop the first Lyapunov-based
adaptive controller built on the KAN architecture, known as Lyapunov
based KAN (Lb-KAN). This work addresses the key challenge of integrating
KANs, which are characterized by learnable activation functions rather
than nested weights with fixed activation functions, into a real-time
control framework that guarantees error convergence and admits visualizable
functional decompositions of system dynamics. This work has three
main contributions. 1) We embed KAN's learnable parameters of activation
functions within a Jacobian matrix in Lyapunov-based adaptation law,
enabling iterative and stable online updates constructed from a Lyapunov-based
analysis. This framework thus establishes an analytic and real-time
learning control process that guarantees asymptotic tracking. 2) We
build a theoretical foundation for KAN-based neural adaptive control
by defining the reconstruction error between the KAN approximation
and the system dynamics based on the KAN approximation theorem {[}15{]}.
3) The Lb-KAN visually describes the functional decomposition of control
system dynamics by representing a complex multivariate function as
a summation of learned univariate functions that is visualizable.
This work serves as a critical foundational step towards future research
focusing on symbolic interpretability in Lyapunov-based adaptive control.

\subsection*{Notation and Preliminaries}

The $n\times n$-dimensional identity matrix is represented by $I_{n}\in\rr^{n\times n}$.
The function composition operator is denoted as $\circ$, i.e., given
functions $f\left(\cdot\right)$ and $g\left(\cdot\right)$, $f\circ g\left(x\right)=f\left(g\left(x\right)\right)$.
The right-to-left matrix product operator is represented by $\stackrel{\curvearrowleft}{\prod}$,
i.e., $\stackrel{\curvearrowleft}{\stackrel[l=1]{m}{\prod}}A_{l}=A_{m}...A_{2}A_{1}$
and $\stackrel{\curvearrowleft}{\stackrel[l=p]{m}{\prod}}A_{p}=1$
if $p>m$. The Kronecker product is denoted by $\otimes$. The vectorization
operator is denoted by $\mathrm{vec}\left(\cdot\right)$, i.e., given
$A\triangleq\left[a_{i,j}\right]\in\mathbb{R}^{m\times n}$, $\mathrm{vec}\left(A\right)\triangleq\left[a_{1,1},\ldots,a_{m,1},\ldots,a_{1,n},\ldots,a_{m,n}\right]{}^{\top}$.
Given any $A\in\rr^{n\times m}$, $B\in\rr^{m\times p}$, and $C\in\rr^{p\times r}$,
$\mathrm{vec}\left(ABC\right)=\left(C^{\top}\otimes A\right)\mathrm{vec}\left(B\right)$,
and $\frac{\partial}{\partial\mathrm{vec}\left(B\right)}\mathrm{vec}\left(ABC\right)=C^{\top}\otimes A.$
The $p$-norm is denoted by $\left\Vert \cdot\right\Vert _{p}$, where
the subscript is suppressed when $p=2$. Almost all time is denoted
as $a.a.t.$ The Filippov set-valued map defined in \cite[Equation 2b]{Paden1987}
is denoted by $K\left[\cdot\right]$. $\left[0,1\right]^{n}$ represents
the $n$-dimensional unit hypercube.

\section{System Dynamics and Control Objective}

Consider a dynamical system modeled as
\begin{align}
\dot{x} & =f\left(x\right)+u+d\left(t\right),\label{eq:Model}
\end{align}
where $x\left(t\right)\in\mathbb{R}^{n}$ denotes the state, $u\left(t\right):\mathbb{R}_{\geq0}\rightarrow\mathbb{R}^{n}$
denotes the control input, and $f\left(x\right):\mathbb{R}^{n}\rightarrow\mathbb{R}^{n}$
denotes an unknown continuously differentiable drift function. The
system disturbances $d\left(t\right)\in\mathbb{R}^{n}$ are assumed
to be bounded as $\left\Vert d\left(t\right)\right\Vert \leq\bar{d}$
where $\bar{d}\in\mathbb{R}_{>0}$ denotes a known constant.

The objective is to design a controller that enables the state $x$
to track a desired trajectory $x_{d}\in\mathbb{R}^{n}$, which is
sufficiently smooth, such that $\left\Vert x_{d}\left(t\right)\right\Vert \le\overline{x_{d}}$
and $\left\Vert \dot{x}_{d}\left(t\right)\right\Vert \le\overline{\dot{x}_{d}}$,
$\forall t\in\mathbb{R}_{\ge0}$, where the constants $\overline{x_{d}},\overline{\dot{x}_{d}}\in\mathbb{R}_{>0}$
are known. 

To evaluate the tracking performance, the tracking error $e\in\mathbb{R}^{n}$
is defined as
\begin{align}
e & \triangleq x-x_{d}.\label{eq: Tracking error}
\end{align}
Taking the time derivative of \eqref{eq: Tracking error} and using
\eqref{eq:Model} yields the open-loop error system
\begin{align}
\dot{e} & =f\left(x\right)+u+d\left(t\right)-\dot{x}_{d}.\label{eq:Pre-Open Loop Dynamics}
\end{align}
The goal is to design a Lb-KAN adaptive controller that ensures $\left\Vert e\left(t\right)\right\Vert \rightarrow0$
as $t\rightarrow\infty$ and provides visualizable functional decomposition
of the unknown dynamics.

\section{Control Development}

\subsection{Lyapunov-Based KAN (Lb-KAN) Architecture}

\begin{figure}[h]
\begin{centering}
\subfloat{

}\includegraphics[scale=0.07]{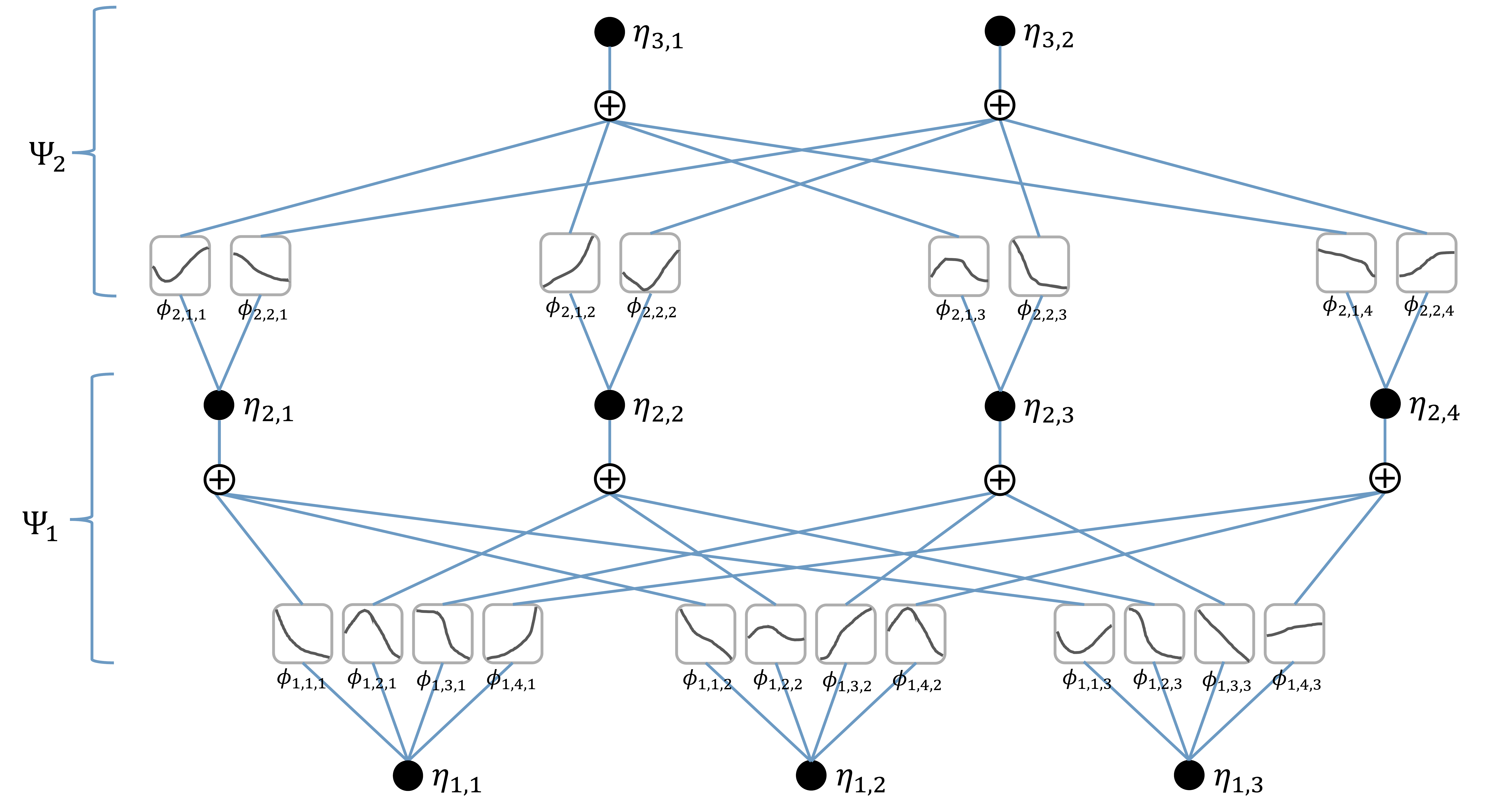}\centering
\par\end{centering}
\caption{\label{fig:KAN pic}An example architecture of a general KAN with
the network shape of $3,4,2$. The blue lines represent the network
edges. Each edge has a learnable univariate activation function. The
simple summation operators in $\Psi_{1}$ and $\Psi_{2}$ are denoted
by $\oplus$.}
\end{figure}

The KART \cite{Kolmogorov1957,Braun2009} states that every multivariate
continuous function $\Phi\left(x\right):\left[0,1\right]^{n}\rightarrow\mathbb{R}$
can be represented as a finite composition of univariate continuous
functions and summation as,
\begin{align}
\Phi\left(\eta\right) & \triangleq\Phi\left(\eta_{1},\ldots,\eta_{n}\right){\color{blue}=}\sum_{q=1}^{2n+1}\Psi_{q}\left(\sum_{p=1}^{n}\phi_{q,p}\left(\eta_{p}\right)\right),\label{eq:KART}
\end{align}
where $\phi_{q,p}:\left[0,1\right]\rightarrow\mathbb{R}$\footnote{The theoretical foundation of KANs stems from the KART, which addresses
function decomposition on the unit domain $\left[0,1\right]^{n}$.
For the Lb- adaptive control problem in this work, the operational
domain is a larger, user-specified compact set $\Upsilon\subset\mathbb{R}^{n}$,
selected to bound the full state trajectory.} and $\Psi_{q}:\mathbb{R}\rightarrow\mathbb{R}$ are the univariate
functions. The architecture in \eqref{eq:KART} can be interpreted
as a shallow, two-layer network with a width of $2n+1$ in the middle
layer. In this specific structure, $\phi_{q,p}$ correspond to learnable
activation functions on edges between input nodes and layer 1 nodes,
$\sum\phi\left(\eta\right)$ represents the summing relation on layer
1 nodes, $\Psi_{q}$ represents learnable activation functions on
edges between layer 1 nodes and layer 2 nodes, and $\sum\Psi_{q}$
represents the summing relation on layer 2 nodes as the output.

A general KAN $\Phi\left(\eta_{1},\theta\right)$ is formulated as
the composition of $L$ layers\footnote{To facilitate the following analysis, it is assumed that activation
functions are smooth, and $\Phi\left(x,\theta\right)$ is assumed
to be continuously differentiable.} \cite{Liu.Wang.ea2024},
\begin{align}
\Phi\left(\eta_{1},\theta\right) & \triangleq\Psi_{L}\circ\Psi_{L-1}\circ\cdots\circ\Psi_{2}\circ\Psi_{1}\left(\eta_{1}\right),\label{eq:KAN}
\end{align}
where $\eta_{1}\in\mathbb{R}^{n}$ denotes the KAN's input, $\theta\triangleq$
$\left[\mathrm{vec}\left(\theta_{1}\right)^{\top}\right.$,$\ldots$,
$\left.\mathrm{vec}\left(\theta_{L}\right)^{\top}\right]^{\top}$
$\in\mathbb{R}^{a_{1}}$ denotes the weights from every layer, $a_{1}\triangleq\left(n_{1}n_{2}+n_{2}n_{3}+...+n_{L}n_{L+1}\right)\left(M+1\right)$,
$M$ is a user-selected constant based on B-spline functions, and
$n_{l}$ denotes the number of nodes in the $l$-th layer. As shown
in Figure \ref{fig:KAN pic}, the KAN's node shape is represented
by an integer array as $\left[n_{1},\ldots,n_{L+1}\right]$, which
specifies the node count of the input layer, hidden layers and output
layer.\footnote{In this paper, the node shape is $n_{1}=n_{L+1}=n$.}
Each $\Psi_{l}\triangleq\left[\eta_{l+1,1},\ldots,\eta_{l+1,n_{l+1}}\right]^{\top}$
$\in\mathbb{R}^{n_{l+1}}$ denotes the activation function summation
in the $l$-th layer for $l\in\left\{ 1,\ldots,L\right\} $. The term
$\eta_{l+1,j}$ is defined as
\begin{align}
\eta_{l+1,j} & \triangleq\sum_{i=1}^{n_{l}}\phi_{l,j,i}\left(\eta_{l,i},\theta_{l,j,i}\right),\label{eq:x_1}
\end{align}
where $\eta_{l+1,j}$ denotes the input on $j$-th node of the $\left(l+1\right)$-th
layer and $j\in\left\{ 1,2,\ldots,n_{l+1}\right\} $, $\eta_{l,i}$
denotes the input on $i$-th node of the $l$-th layer where \textbf{$i\in\left\{ 1,2,\ldots,n_{l}\right\} $},
and $\theta_{l,j,i}$ is the parameters in the activation function
$\phi_{l,j,i}$. Between layer $l$ and layer $l+1$, there are $n_{l}n_{l+1}$
activation functions on the edges, and $\phi_{l,j,i}\left(\eta_{l,i},\theta_{l,j,i}\right)$
denotes the activation function from $n_{l}$ nodes to $n_{l+1}$
nodes, which is defined as the sum of the basis function $b\left(\eta_{l,i}\right)$
and the B-spline function $s\left(\eta_{l,i}\right)$
\begin{align}
\phi_{l,j,i}\left(\eta_{l,i},\theta_{l,j,i}\right) & \triangleq w_{b,l,j,i}b\left(\eta_{l,i}\right)+w_{s,l,j,i}s\left(\eta_{l,i}\right),\nonumber \\
 & =\left[w_{b,l,j,i}\,\,w_{s,l,j,i}\right]\left[b\left(\eta_{l,i}\right)\,\,s\left(\eta_{l,i}\right)\right]^{\top},\label{eq:activation function definition}
\end{align}
where $b\left(\eta_{l,i}\right)\triangleq\frac{\eta_{l,i}}{1+e^{-\eta_{l,i}}}$,
$s\left(\eta_{l,i}\right)\triangleq\sum_{m=1}^{M}c_{m,l,j,i}B_{m}\left(\eta_{l,i}\right)$,
$M\triangleq G+k$ where $k$ is the order of B-spline and $G$ is
the grid size, $w_{b,l,j,i}$, $w_{s,l,j,i}$ and $c_{m,l,j,i}$ are
the parameters to adjust the overall magnitude of the activation function,
and $\theta_{l,j,i}\triangleq$ $\left[w_{b,l,j,i}\right.$, $w_{s,l,j,i}c_{1,l,j,i}$,
$w_{s,l,j,i}c_{2,l,j,i}$,$\ldots$, $\left.w_{s,l,j,i}c_{M,l,j,i}\right]$
$\in\mathbb{R}^{1\times(M+1)}$. Define $X_{l,i}\triangleq$ $\left[b\left(\eta_{l,i}\right)\right.$,
$B_{1}\left(\eta_{l,i}\right)$, $B_{2}\left(\eta_{l,i}\right)$,$\ldots$,
$\left.B_{M}\left(\eta_{l,i}\right)\right]^{\top}$ $\in\mathbb{R}^{M+1}$.
Then, each activation function can be expressed as $\phi_{l,j,i}\left(\eta_{l,i},\theta_{l,j,i}\right)=\theta_{l,j,i}X_{l,i}$.
As a result, from \eqref{eq:x_1} we have

\begin{align}
\Psi_{l} & \triangleq\left[\begin{array}{c}
\eta_{l+1,1}\\
\eta_{l+1,2}\\
\vdots\\
\eta_{l+1,n_{l+1}}
\end{array}\right],\nonumber \\
 & \triangleq\underbrace{\left[\begin{array}{ccc}
\theta_{l,1,1} & \cdots & \theta_{l,1,n_{l}}\\
\theta_{l,2,1} & \cdots & \theta_{l,2,n_{l}}\\
\vdots &  & \vdots\\
\theta_{l,n_{l+1},1} & \cdots & \theta_{l,n_{l+1},n_{l}}
\end{array}\right]}_{\theta_{l}}\left[\begin{array}{c}
X_{l,1}\\
X_{l,2}\\
\vdots\\
X_{l,n_{l}}
\end{array}\right],\label{eq:theta matrix}
\end{align}
where $\theta_{l}\in\mathbb{R}^{n_{l+1}\times n_{l}(M+1)}$.
\begin{assumption}
\label{assm:aup1}There exists a known constant $\bar{\theta}\in\rr_{>0}$
such that the ideal weights $\theta$ in \eqref{eq:KAN} can be bounded
as $\left\Vert \theta\right\Vert \leq\bar{\theta}$.\footnote{The robust adaptive work in results such as \cite{Stepanyan.Kurdila2009}
could provide extensions for an unknown $\bar{\theta}$.}

Based on the approximation theorem of KAN \cite[Theorem 2.1]{Liu.Wang.ea2024},
let $\mathcal{C}\left(\mathcal{\mathcal{X}}\right)$ represent the
space of continuous functions over the set $\mathcal{X}\subseteq\rr^{n}$,
where $x\in\mathcal{X}$. The function space of KAN is dense in $\mathcal{C}\left(\mathcal{X}\right)$.
Then, the unknown drift dynamics $f\left(x\right)$ in \eqref{eq:Model}
can be modeled by the KAN architecture in \eqref{eq:KAN} as
\begin{align}
f\left(x\right) & =\Phi\left(x,\theta\right)+\varepsilon,\label{eq:reconstruction error}
\end{align}
where $\varepsilon\in\mathbb{R}^{n}$ represents an unknown function
reconstruction error and $\eta_{1}=x$. From Assumption \ref{assm:aup1},
for any given $\overline{\varepsilon}\in\rr_{>0}$, there exists an
ideal weight matrix $\theta$ such that $\left\Vert \varepsilon\right\Vert \leq\overline{\varepsilon},\forall x\in\mathcal{X}$,
where $\overline{\varepsilon}$ is partially known. From \cite[Theorem 2.1]{Liu.Wang.ea2024},
the $\mathbb{C}^{m}$-norm bound is given by $\overline{\varepsilon}\triangleq CG^{-k-1+m}$
where $m$ represents the highest order of the function's derivative
being approximated where $0\leq m\leq k$, the grid size $G$ and
the order of B-spline $k$ are user-defined before training, and the
constant $C$ is independent of $G$. Although the independent constant
$C$ is influenced by $\Phi\left(x\right)$ and its representation,
it would still exist due to the continuity of $f$ and $\Phi$. In
our case, $m=0$. Thus, $\overline{\varepsilon}$ is independent of
the input dimension, implying that KANs do not suffer from the curse
of dimensionality \cite{Liu.Wang.ea2024}. This differs from MLP architectures
that are based on the UAT where $\overline{\varepsilon}$ is dependent
on the number of neurons which often scales exponentially with the
input dimension, leading to the curse of dimensionality.
\end{assumption}

\subsection{Lb-KAN Weight Adaptation Law}

Using the KAN model in \eqref{eq:KAN}, an adaptive KAN estimate using
the shorthand notation $\widehat{\Phi}\triangleq\Phi\left(x,\hat{\theta}\right)$$\in\mathbb{R}^{n}$
is constructed to approximate the unknown drift dynamics $f\left(x\right)$
in \eqref{eq:Model} via the weight estimate. The overall weight estimate
$\hat{\theta}\in\mathbb{R}^{a_{1}}$ is defined as $\hat{\theta}\triangleq\left[\begin{array}{ccc}
\mathrm{vec}\left(\hat{\theta}_{1}\right)^{\top} & ,\ldots, & \mathrm{vec}\left(\hat{\theta}_{L}\right)^{\top}\end{array}\right]^{\top}$, \textbf{$\hat{\theta}_{l}\in\mathbb{R}^{n_{l+1}\times n_{l}(M+1)}$}
denotes the estimates of $\theta_{l}$ in \eqref{eq:theta matrix},
and $\hat{\theta}_{l,j,i}\triangleq$$\left[\hat{w}_{b,l,j,i}\right.$,
$\hat{w}_{s,l,j,i}\hat{c}_{1,l,j,i}$, $\hat{w}_{s,l,j,i}\hat{c}_{2,l,j,i}$,
$\ldots$, $\left.\hat{w}_{s,l,j,i}\hat{c}_{M,l,j,i}\right]\in\mathbb{R}^{1\times(M+1)}$.
Based on \eqref{eq:activation function definition}, the activation
function estimate can be defined as $\hat{\phi}_{l,j,i}\left(x_{l,i},\hat{\theta}_{l,j,i}\right)\triangleq\hat{\theta}_{l,j,i}X_{l,i}$,
$i\in\left\{ 1,2,\ldots,n_{l}\right\} $, $j\in\left\{ 1,2,\ldots,n_{l+1}\right\} $.
As a result, $\widehat{\Psi}_{l}$ is the matrix of these activation
function estimates $\hat{\phi}_{l,j,i}$ as 
\begin{align*}
\widehat{\Psi}_{l} & =\left[\begin{array}{c}
\sum_{i=1}^{n_{l}}\hat{\phi}_{l,1,i}\left(\eta_{l,i},\hat{\theta}_{l,1,i}\right)\\
\sum_{i=1}^{n_{l}}\hat{\phi}_{l,2,i}\left(\eta_{l,i},\hat{\theta}_{l,2,i}\right)\\
\vdots\\
\sum_{i=1}^{n_{l}}\hat{\phi}_{l,n_{l+1},i}\left(\eta_{l,i},\hat{\theta}_{l,n_{l+1},i}\right)
\end{array}\right].
\end{align*}

The adaptive weight estimate $\hat{\theta}$ is implemented and updated
by a Lyapunov-based weight adaptation law (i.e., the adaptive update
law is constructed based on insights from the Lyapunov-based analysis)
\begin{align}
\dot{\hat{\theta}} & \triangleq\text{proj}\left(\Gamma\widehat{\Phi}^{\prime\top}e\right),\label{eq:Weight Update Law}
\end{align}
where $\Gamma\in\mathbb{R}^{a_{1}\times a_{1}}$ is a positive-definite
adaptation gain matrix, the projection operator ensures $\hat{\theta}\left(t\right)\in\bar{\hat{\theta}}$
for all $t\in\rr_{>0}$ \cite[Appendix E]{Krstic.Kanellakopoulos.ea1995},
and $\widehat{\Phi}^{\prime}$ is a short-hand notation denoting the
Jacobian $\widehat{\Phi}^{\prime}\triangleq\frac{\partial\widehat{\Phi}}{\partial\widehat{\theta}}\triangleq\left[\widehat{\Psi}_{1}^{\prime},\ldots,\widehat{\Psi}_{L}^{\prime}\right]$
$\in\mathbb{R}^{n\times a_{1}}$. 

To facilitate the subsequent development, let the shorthand notation
$\widehat{\Psi}_{l}^{\prime}$ be defined as $\widehat{\Psi}_{l}^{\prime}\triangleq\frac{\partial\widehat{\Phi}}{\partial\mathrm{vec}\left(\hat{\theta}_{l}\right)}$,
$l\in\left\{ 1,\ldots,L\right\} $. Taking the partial derivative
of $\widehat{\Phi}$, the term $\widehat{\Psi}_{l}^{\prime}$ is expressed
in matrix form as
\begin{align}
\widehat{\Psi}_{l}^{\prime} & \triangleq\left(\stackrel{\curvearrowleft}{\stackrel[v=l+1]{L}{\prod}}\Xi_{v}\right)\Lambda_{l},\:\:\forall l\in\left\{ 1,\ldots,L\right\} ,\label{eq:jacobian}
\end{align}
where $\Lambda_{l}\triangleq\frac{\partial\widehat{\Psi}_{l}}{\partial vec\left(\widehat{\theta}_{l}\right)}$
and $\Xi_{l}\triangleq\frac{\partial\widehat{\Psi}_{l}}{\partial x_{l}}$.
Using the properties of the vectorization operator, the terms in $\Lambda_{l}$
can be computed as
\begin{align*}
\Lambda_{l} & =X_{l}^{\top}\otimes I_{n_{l+1}},
\end{align*}
where $X_{l}\triangleq\left[\begin{array}{cccc}
X_{l,1}^{\top} & X_{l,2}^{\top} & \cdots & X_{l,n_{l}}^{\top}\end{array}\right]^{\top}\in\mathbb{R}^{n_{l}(M+1)}$ and $X_{l,i}\triangleq$$\left[b\left(x_{l,i}\right)\right.$, $B_{1}\left(x_{l,i}\right)$,
$B_{2}\left(x_{l,i}\right)$, $\ldots$, $\left.B_{M}\left(x_{l,i}\right)\right]^{\top}\in\mathbb{R}^{M+1}$,
$i\in\left\{ 1,2,\ldots,n_{L}\right\} $. Using the chain rule, the
terms in $\Xi_{l}$ can be computed as
\begin{align*}
\Xi_{l} & =\widehat{\theta}_{l}X_{l}^{\prime},
\end{align*}
where $X_{l}^{\prime}=\frac{\partial X_{l}}{\partial x_{l}}$, which
includes the derivative of the known basis function $b^{\prime}\left(x\right)$
and the derivative of the B-spline $s^{\prime}\left(x\right)$. Given
$k,G$ and input range $\left[-c,c\right]$, $s^{\prime}\left(x\right)$
can be computed by the B-spline differentiation formula \cite{De1978}.

Inspired by Lyapunov-based NN control methods based on MLP architectures
in \cite{Patil.Le.ea2022} and \cite{Griffis.Patil.ea23_2}, we use
a first order Taylor series approximation \cite[Eq. 22]{Lewis1996b}
of the estimation error to develop the Lb-KAN as
\begin{align}
\widetilde{\Phi} & \triangleq\widehat{\Phi}^{\prime}\tilde{\theta}+R\left(x,\left\Vert \tilde{\theta}\right\Vert ^{2}\right),\label{eq:Taylor Series}
\end{align}
where $\tilde{\theta}\triangleq\theta-\hat{\theta}\in\mathbb{R}^{a_{1}}$
denotes the weight estimation error, the shorthand notation $\widetilde{\Phi}$
is defined as $\widetilde{\Phi}\triangleq\Phi-\widehat{\Phi}\in\mathbb{R}^{n}$,
and the term $R\left(x,\left\Vert \tilde{\theta}\right\Vert ^{2}\right)\in\mathbb{R}^{n}$
denotes the first Lagrange remainder term.

\subsection{Controller Design}

Based on the open-loop error system in \eqref{eq:Pre-Open Loop Dynamics},
and the subsequent stability analysis, the control input $u:\mathbb{R}_{\geq0}\rightarrow\mathbb{R}^{n}$
is designed as
\begin{align}
u & \triangleq-\widehat{\Phi}-k_{e}e-k_{s}\text{sgn\ensuremath{\left(e\right)}}+\dot{x}_{d},\label{eq:Control law}
\end{align}
where $k_{e},k_{s}\in\rr_{>0}$ denote user-selected constants, and
the term $k_{s}\text{sgn\ensuremath{\left(e\right)}}$ is a sliding
mode control and designed to compensate for residual system uncertainties.\footnote{The KAN estimate is developed to approximate the uncertainty in unknown
dynamical system $f\left(x\right)$, but additional robust terms are
required to compensate for residual disturbances such as the residual
function approximation error $\varepsilon$ in \eqref{eq:reconstruction error}
and the residual terms $R\left(x,\left\Vert \tilde{\theta}\right\Vert ^{2}\right)$.
High gain state feedback can compensate for these residual terms at
the expense of a yielding a uniformly ultimately bounded result. Likewise,
various other robust control design approaches could also be used.
In this paper, we elected to use sliding mode control to compensate
for residual terms to yield an asymptotic convergence result.} Substituting \eqref{eq:reconstruction error}, \eqref{eq:Taylor Series}
and \eqref{eq:Control law} into \eqref{eq:Pre-Open Loop Dynamics},
the resulting closed-loop error system is
\begin{align}
\dot{e} & =\widehat{\Phi}^{\prime}\tilde{\theta}+R+\varepsilon+d\left(t\right)-k_{e}e-k_{s}\text{sgn\ensuremath{\left(e\right)}}.\label{eq: Closed-Loop Error System}
\end{align}

\section{\label{sec:Stability-Analysis}Stability Analysis}

In this section, a Lyapunov-based stability analysis is provided for
the developed controller in \eqref{eq:Control law} with the Lb-KAN
update law from \eqref{eq:Weight Update Law}. To do this, consider
the Lyapunov function candidate $\mathcal{V}_{L}:\mathbb{R}^{a_{2}}\rightarrow\mathbb{R}_{\geq0}$
defined as
\begin{align}
\mathcal{V}_{L}\left(z\right) & \triangleq\frac{1}{2}e^{\top}e+\frac{1}{2}\tilde{\theta}^{\top}\Gamma^{^{-1}}\tilde{\theta},\label{eq:Lyapunov function}
\end{align}
where $z\triangleq\left[\begin{array}{cc}
e^{\top}, & \tilde{\theta}^{\top}\end{array}\right]^{\top}\in\mathbb{R}^{a_{2}}$ and $a_{2}\triangleq n+a_{1}$. The candidate Lyapunov function in
\eqref{eq:Lyapunov function} satisfies $\beta_{1}\left\Vert z\right\Vert ^{2}\leq\mathcal{V}_{L}\left(z\right)\leq\beta_{2}\left\Vert z\right\Vert ^{2}$,
where $\beta_{1}\triangleq\text{min}\left\{ \frac{1}{2},\frac{1}{2}\lambda_{\text{min}}\left\{ \Gamma^{^{-1}}\right\} \right\} $
and $\beta_{2}\triangleq\text{max}\left\{ \frac{1}{2},\frac{1}{2}\lambda_{\text{max}}\left\{ \Gamma^{^{-1}}\right\} \right\} $. 

Since the signum term is discontinuous, a generalized time-derivative
is used that is based on a Filippov set-valued map. Taking the generalized
time derivative of $\mathcal{V}_{L}$ yields
\begin{align}
\dot{\mathcal{V}}_{L} & \overset{a.a.t.}{\in}e^{\top}K\text{\ensuremath{\left[\dot{e}\right]}}-\tilde{\theta}^{\top}\Gamma^{-1}\dot{\hat{\theta}}.\label{eq:Lyapunov 1}
\end{align}
Substituting the weight adaptation law in \eqref{eq:Weight Update Law}
and the closed-loop error system in \eqref{eq: Closed-Loop Error System}
into \eqref{eq:Lyapunov 1} yields
\begin{align*}
\dot{\mathcal{V}}_{L} & \overset{a.a.t.}{\in}e^{\top}\left(\widehat{\Phi}^{\prime}\tilde{\theta}+R+\varepsilon+d\left(t\right)-k_{e}e-k_{s}\text{sgn\ensuremath{\left(e\right)}}\right)\\
 & \qquad-\tilde{\theta}^{\top}\Gamma^{-1}\left(\Gamma\widehat{\Phi}^{\prime\top}e\right).
\end{align*}
Applying $K\left[\cdot\right]$ and canceling out the last term yields
\begin{align}
\dot{\mathcal{V}}_{L} & \overset{a.a.t.}{\leq}-k_{e}e^{\top}e-k_{s}\left\Vert e\right\Vert _{1}+e^{\top}\left(\varepsilon+R+d\left(t\right)\right).\label{eq:Lyapunov 2}
\end{align}

Let the open and connected compact sets $\mathcal{H}\subset\rr^{a_{2}}$
and $\mathcal{S}\subset\rr^{a_{2}}$ be defined as $\mathcal{H}\triangleq\left\{ \varsigma\in\rr^{a_{2}}:\left\Vert \varsigma\right\Vert \leq\sqrt{\frac{\beta_{1}}{\beta_{2}}}\omega\right\} $
and $\mathcal{S}\triangleq\left\{ \varsigma\in\rr^{a_{2}}:\lVert\varsigma\rVert\leq\omega\right\} $,
respectively, where $\omega\in\rr_{>0}$ denotes a user-selected bounding
constant. To facilitate the analysis, the upper bounds for the uncertainty
terms $\varepsilon$, $R$, and $d$ are established as follows. Since
$\left\Vert x_{d}\right\Vert \leq\bar{x}_{d}$, $\forall t\in\mathbb{R}_{\ge0}$,
$x$ can be bounded as $\left\Vert x\right\Vert \leq\left\Vert e+x_{d}\right\Vert \leq\left\Vert z\right\Vert +\left\Vert x_{d}\right\Vert \leq\omega+\bar{x}_{d}$
when $z\in\mathcal{S}$. The term $\tilde{\theta}$ is bounded since
$\theta$ is a constant and $\hat{\theta}$ is inside a projection
algorithm. Thus, there is $\left\Vert R\left(x,\left\Vert \tilde{\theta}\right\Vert ^{2}\right)\right\Vert \leq\overline{R}$
where $\overline{R}\in\mathbb{R}_{>0}$ when $z\in\mathcal{S}$. Since
$\left\Vert \varepsilon\right\Vert \leq\overline{\varepsilon}$ and
$\left\Vert d\left(t\right)\right\Vert \leq\overline{d}$, substituting
these bounds into \eqref{eq:Lyapunov 2} yields
\begin{align}
\dot{\mathcal{V}}_{L} & \overset{a.a.t.}{\leq}-k_{e}e^{\top}e-\left\Vert e\right\Vert _{1}\left(k_{s}-\overline{\varepsilon}-\overline{R}-\overline{d}\right),\label{eq:Lyapunov 3}
\end{align}
when $z\in\mathcal{\mathcal{S}}$. Since the approximation theory
in \eqref{eq:KART} is on the compact set $\mathcal{X}$, it is shown
in the following analysis that $x\in\mathcal{X}$ $\forall t\geq0$
by proving $z\in\mathcal{S}$, $\forall t\geq0$ when $z\left(0\right)\in\mathcal{H}$.
Then, it can be shown that $x\in\mathcal{X}$, $\forall t\geq0$,
and therefore, the approximation theory of KART holds.
\begin{thm}
The controller in \eqref{eq:Control law} and the weight adaptation
law in \eqref{eq:Weight Update Law} ensure asymptotic tracking error
convergence in the sense that $\underset{t\rightarrow\infty}{\text{lim}}\left\Vert e\left(t\right)\right\Vert =0$,
provided $z\left(0\right)\in\mathcal{H}$ and the following gain condition
is satisfied
\begin{align}
k_{s} & >\overline{\varepsilon}+\overline{R}+\bar{d}.\label{eq:Gainconditions}
\end{align}
\end{thm}
\begin{IEEEproof}
Consider the candidate Lyapunov function in \eqref{eq:Lyapunov function}.
Provided the sufficient gain condition in \eqref{eq:Gainconditions}
is met, \eqref{eq:Lyapunov 3} can be bounded as
\begin{align}
\dot{\mathcal{V}}_{L} & \overset{a.a.t.}{\leq}-k_{e}e^{\top}e,\label{eq:Lyapunov 4}
\end{align}
when $z\in\mathcal{S}$. It remains to be shown that $z\in\mathcal{S}$,
$\forall t\geq0$. Using \eqref{eq:Lyapunov function} and the fact
that $\dot{\mathcal{V}}_{L}\left(z\left(t\right)\right)\overset{a.a.t.}{\leq}0$
when $z\in\mathcal{S}$, we have $\beta_{1}\left\Vert z\left(t\right)\right\Vert ^{2}\leq\mathcal{V}_{L}\left(z\left(t\right)\right)$
$\leq\mathcal{V}_{L}\left(z\left(0\right)\right)\leq\beta_{2}\left\Vert z\left(0\right)\right\Vert ^{2}$.
Thus, $\lVert z\left(t\right)\rVert\leq\sqrt{\frac{\beta_{2}}{\beta_{1}}}\lVert z\left(0\right)\rVert$
when $z\in\mathcal{S}$. If $\lVert z\left(0\right)\rVert\leq\omega\sqrt{\frac{\beta_{1}}{\beta_{2}}},$
then $\lVert z\left(t\right)\rVert\leq\omega$, $\forall t\geq0$.
Therefore, if $z$ is initialized such that $z\left(0\right)\in\mathcal{H}$,
then $z\in\mathcal{S}$, $\forall t\geq0$. It remains to be shown
that $x\in\mathcal{X}$. Let the open and connected set $\Upsilon\subseteq\mathcal{X}$
be defined as $\Upsilon\triangleq\left\{ \varsigma\in\mathcal{X}:\lVert\varsigma\rVert\leq\omega+\overline{x}_{d}\right\} $.
Since $\lVert z\left(t\right)\rVert\leq\omega$, $\forall t\geq0$,
then $\lVert e\left(t\right)\rVert\leq\omega$, $\forall t\geq0$.
Hence, using \eqref{eq: Tracking error}, $x$ can be bounded as $\left\Vert x\right\Vert \leq\omega+\overline{x}_{d}$.
Therefore, if $z\left(0\right)\in\mathcal{H}$, then $x\in\Upsilon\subseteq\mathcal{\mathcal{X}}$.
Using \eqref{eq:Lyapunov function} and $\dot{\mathcal{V}}_{L}\overset{a.a.t.}{\leq}0$,
we have $e,\tilde{\theta}\in\mathcal{L}_{\infty}$. Using $e,\tilde{\theta},\theta,x_{d}\in\mathcal{L}_{\infty}$
implies $x,\hat{\theta}\in\mathcal{L}_{\infty}$. Using \eqref{eq:Weight Update Law},
the fact that $x,\hat{\theta}\in\mathcal{L}_{\infty}$ and the fact
that $\widehat{\Phi}$ is smooth implies $\dot{\hat{\theta}}\in\mathcal{L}_{\infty}$.
Using \eqref{eq:Control law} and the fact that $x,e\in\mathcal{L}_{\infty}$
implies $u\in\mathcal{L}_{\infty}$. Then, using \eqref{eq:Lyapunov 4}
and the extension of the LaSalle-Yoshizawa theorem for non-smooth
systems \cite[Corollary 1, Corollary 2]{Fischer.Kamalapurkar.ea2013},
it can be shown that $\underset{t\rightarrow\infty}{\text{lim}}\left\Vert e\left(t\right)\right\Vert =0$
when $z\in\mathcal{S}$, resulting in asymptotic tracking error convergence.
\end{IEEEproof}

\section{Simulations}

\subsection{Tracking and Approximation Performance}

\begin{figure}[h]
\begin{centering}
\includegraphics[scale=0.25]{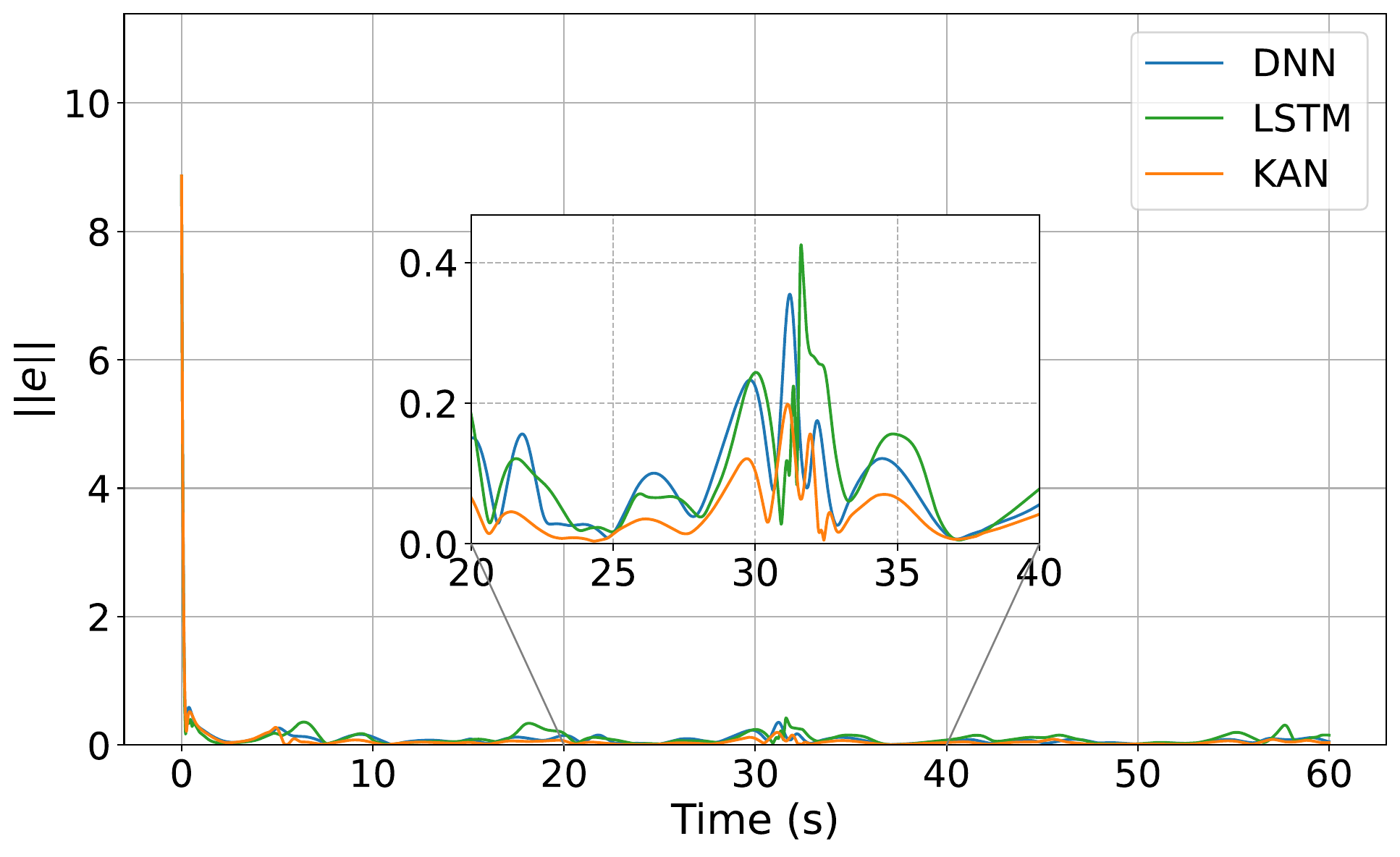}\centering
\par\end{centering}
\caption{\label{fig:Error Plots}Plots of the norms of RMS tracking error $\left\Vert e\right\Vert $
over time \textcolor{blue}{for} Lb-KAN, Lb-LSTM, and Lb-DNN adaptive
controllers for one representative run.}
\end{figure}
\begin{figure}[h]
\begin{centering}
\includegraphics[scale=0.25]{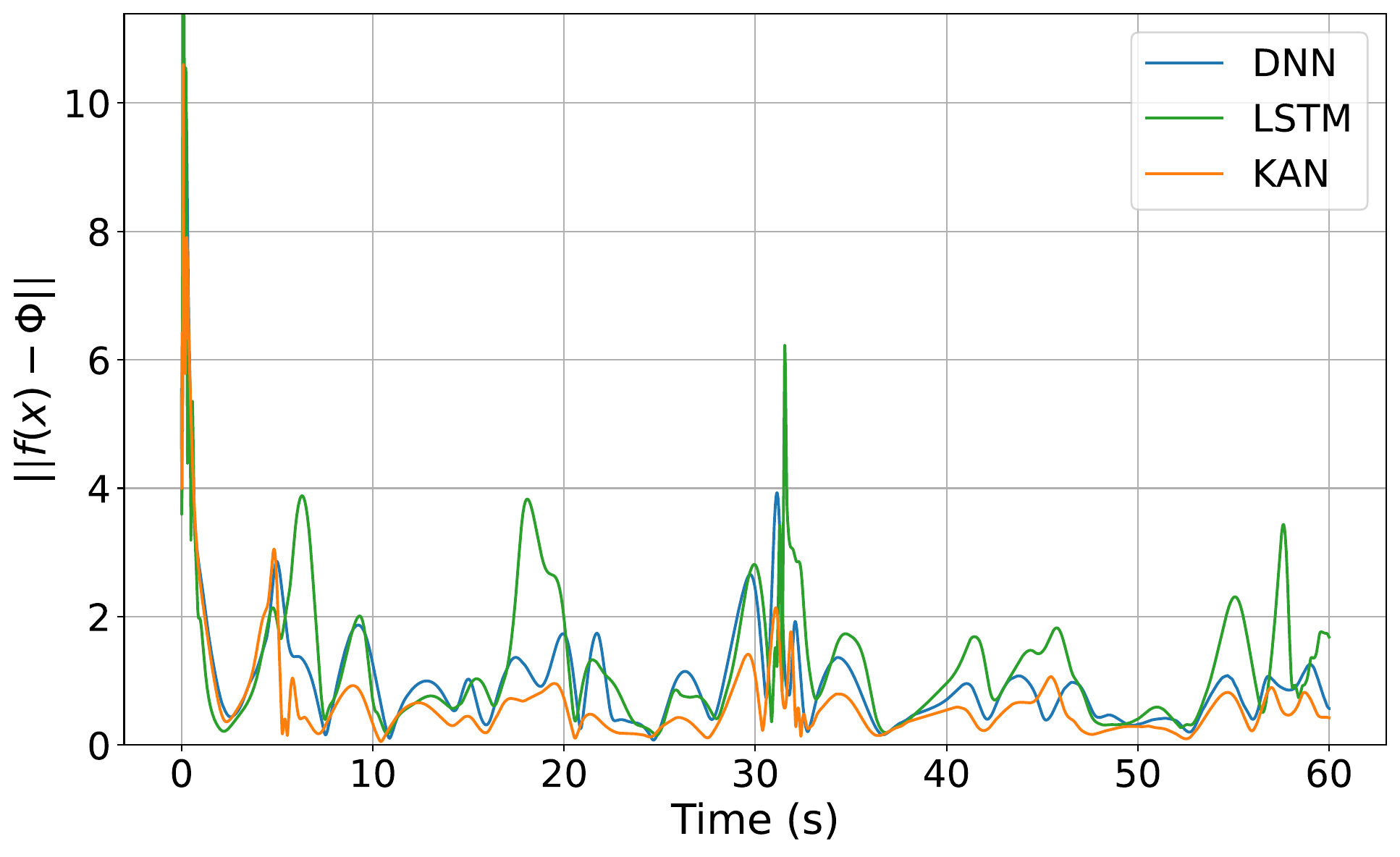}\centering
\par\end{centering}
\caption{\label{fig:ft Plots}Plots of the norms of RMS function approximation
error $\left\Vert f\left(x\right)-\widehat{\Phi}\right\Vert $ over
time for Lb-KAN, Lb-LSTM, and Lb-DNN adaptive controllers for one
representative run.}
\end{figure}
The Lb-KAN adaptive controller is verified in simulations on a four-state
nonlinear dynamical system. For comparison, two baseline methods are
included: the Lb-DNN in \cite{Patil.Le.ea2022} and the Lb-LSTM in
\cite{Griffis.Patil.ea23_2} for the same dynamical system.

The drift dynamics in \eqref{eq:Model} are modeled as $f\left(x\right)\triangleq$
$\left[4\text{tanh}\left(x_{1}+\text{sin}\left(\pi x_{2}\right)\right)\right.$,
$5e^{-\left(x_{2}^{2}+x_{3}^{2}\right)}-2$, $3\text{sin}\left(\pi\left(x_{1}+x_{4}\right)\right)+2\text{cos\ensuremath{\left(\pi\left(x_{3}+x_{2}\right)\right)}}$,
$\left.\frac{4}{1+e^{-\left(x_{1}-x_{2}\right)}}+\text{sin\ensuremath{\left(2x_{4}\right)}-2}\right]^{\top}$,
which is unknown. The disturbance is given by $d\left(t\right)=0.1\text{cos}\left(0.5t\right)$.
The simulations are performed for 100s with a 0.001s step size. The
initial weights are generated from a uniform distribution $U\left(-0.1,0.1\right)$.
The initial state is $x\left(0\right)\sim U\left(-8,8\right)$. The
desired trajectory is designed as $x_{d}\left(t\right)=\left[a\text{sin}\left(bt+c\right),\ldots,a\text{sin}\left(bt+c\right)\right]^{\top}$,
where $a\sim U\left(0.5,1.5\right)$ rad, $b\sim U\left(0.2,0.5\right)$
rad/s and $c\sim U\left(0,\pi\right)$ rad. The constant for projection
operator is set as $\bar{\hat{\theta}}=5$.

The LSTM and DNN controllers are constructed by replacing the Lb-KAN
term in \eqref{eq:Control law} with the adaptive Lb-LSTM model in
\cite{Griffis.Patil.ea23_2} and the adaptive Lb-DNN model in \cite{Patil.Le.ea2022},
respectively. To ensure a fair comparison, the controller gains are
selected as $k_{e}=11$ and $k_{s}=0.01$, which are applied for all
three controllers. However, to allow each architecture to achieve
its minimal function approximation errors, the adaptation gain matrix
$\Gamma$ is tuned individually, where $\Gamma$ is selected as $6.0I_{576}$,
$20.0I_{576}$, and $4.2I_{576}$ for the DNN, LSTM, and KAN controllers,
respectively.

The width of Lb-DNN is set as $\left[4,5,5,5,4\right]$ and the activation
function is tanh. The Lb-LSTM model is implemented with five neurons,
meaning both its internal cell state and hidden state are five-dimensional
vectors. The width of Lb-KAN architecture is set as $\left[4,6,4,4\right]$,
the grid size $G=5$ and the B-spline order $k=3$.

Since the online adaptation task for learning models are sensitive
to weight initialization, a Monte Carlo approach is used to initialize
the weight estimates. In this method, $1000$ simulations are performed,
where the initial weights in each simulation are selected from $U(-0.1,0.1)$,
and the cost $J=\int_{0}^{60}\left(f\left(x\right)-\widehat{\Phi}\left(t\right)\right)^{\top}\left(f\left(x\right)-\widehat{\Phi}\left(t\right)\right)dt$
is evaluated in each simulation. The weights with the lowest cost
are selected as the initial conditions.

As shown in Figure \ref{fig:Error Plots}, the developed Lb-KAN controller
demonstrates robust trajectory tracking performance comparable to
the Lb-DNN and Lb-LSTM benchmarks. All three controllers achieve rapid
convergence of the tracking error within approximately 0.5 seconds
and effectively maintain the state trajectories around the desired
trajectories despite time-varying disturbances.

\begin{table}[tbh]
\caption{\label{tab:tableKAN}Mean and Standard Deviation of Performance Comparison
Results for 20 runs}

\centering{}%
\begin{tabular}{|c|c|c|}
\hline 
Network Architectures & \multirow{1}{*}{$\left\Vert e\right\Vert $} & $\left\Vert f\left(x\right)-\widehat{\Phi}\right\Vert $\tabularnewline
\hline 
KAN & 0.288$\pm$0.0169 & 1.04$\pm$0.139\tabularnewline
\hline 
LSTM & 0.292$\pm$0.0176 & 1.31$\pm$0.195\tabularnewline
\hline 
DNN & 0.293$\pm$0.0167 & 1.27$\pm$0.189\tabularnewline
\hline 
\end{tabular}
\end{table}
While the tracking accuracy is similar across the three architectures,
the Lb-KAN shows higher fidelity in approximating the unknown system
dynamics, as shown in Figure \ref{fig:ft Plots}. Table \ref{tab:tableKAN}
presents the mean and standard deviation of performance for 20 runs
with randomly generated desired trajectories. The advantage is prominent
in function approximation, where the Lb-KAN outperforms the Lb-LSTM
and Lb-DNN by 20.2\% and 18.0\%, respectively.

\subsection{Visualization of Explicit Functional Decomposition}

\begin{figure}[h]
\begin{centering}
\includegraphics[scale=0.065]{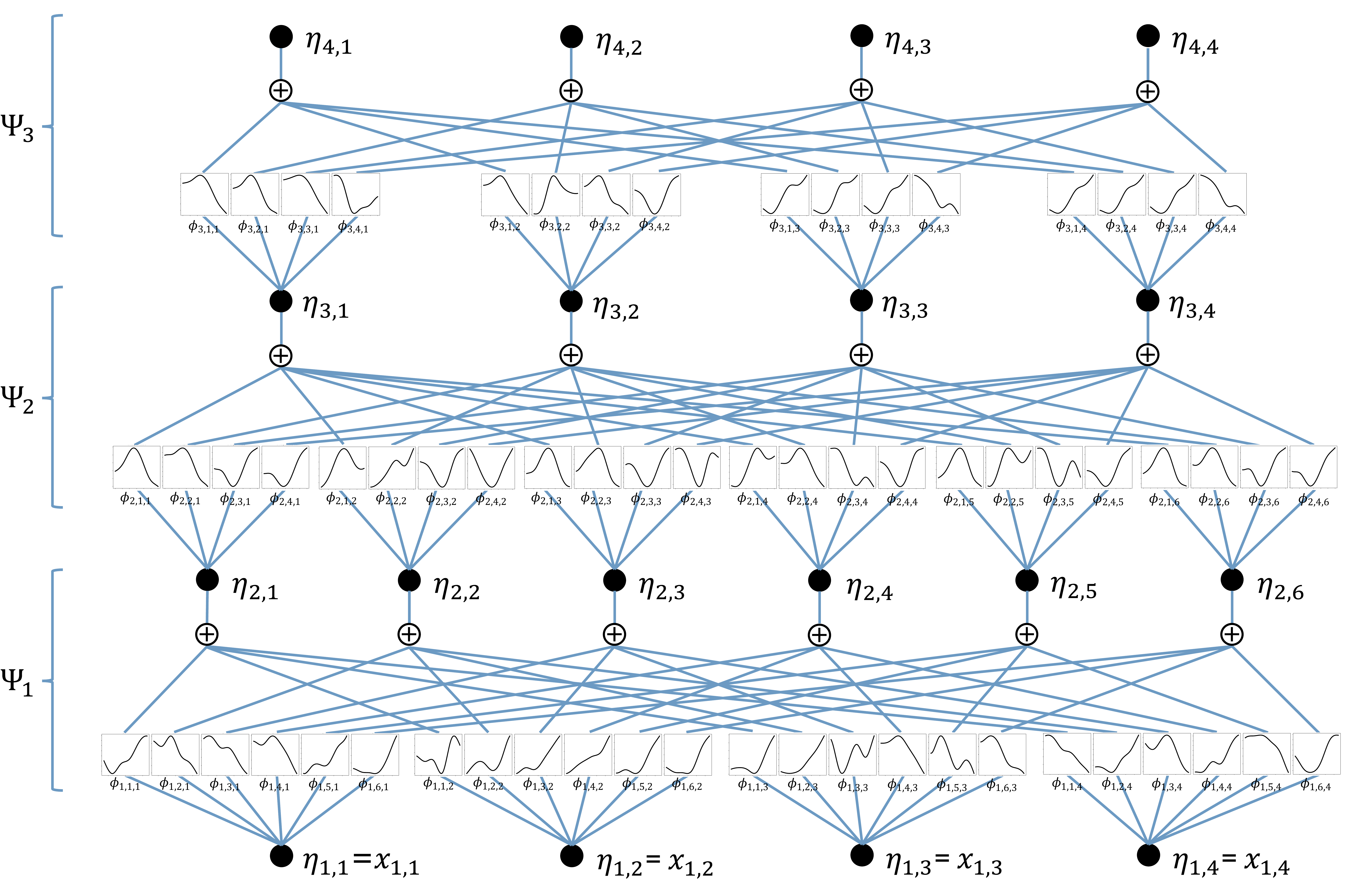}\centering
\par\end{centering}
\caption{\label{fig:int Plot}Internal representations in KAN.}
\end{figure}
In addition to tracking and approximation performance, a core architectural
feature of the Lb-KAN is its capacity for explicit functional decomposition,
which represents the unknown system dynamics as a summation of univariate
functions. As shown in Figure \ref{fig:int Plot}, this structure
permits direct visual inspection of the learned functions.

However, explicit functional decomposition is not equivalent to symbolic
interpretability. While functional decomposition offers visualization,
symbolic interpretability implies that the exact mathematical components
of the dynamics are uniquely recovered. Achieving symbolic interpretability
requires a sparse network structure, where only the nodes uniquely
corresponding to the true dynamic components are active.

To achieve symbolically interpretable adaptive control, the sparsity-inducing
terms (e.g., L1 and entropy regularization) would need to be integrated
into the design of the adaptation law. While our Lb-KAN achieves higher
fidelity in approximating the unknown system dynamics, the learned
functions in Figure \ref{fig:int Plot} exhibit a distributed representation
rather than a sparse representation. Each activation function contributes
to the approximation, resulting in a network characterized by high
entropy. This signifies that while the system dynamics functions are
explicitly decomposed, they are not yet condensed into the concise
symbolic equations characterizing true symbolic interpretability.

Therefore, this work serves as a foundational stepping stone for future
study on symbolically interpretable adaptive control. Having established
performance of the Lb-KAN framework, our future research aims to bridge
this gap, developing a real-time adaptive control method with symbolic
interpretability, guaranteed convergence and tracking performance.

\section{Conclusions}

This paper presents the first Lyapunov-based adaptive control framework
utilizing KANs, referred to as Lb-KANs, to address the challenges
of real-time learning and convergence in tracking control of uncertain
nonlinear systems. By embedding KANs into a Lyapunov-based adaptive
control framework, the developed approach enables explicit functional
decomposition and better function approximation with online updates.
The controller integrates learnable activation parameters into the
adaptation law, enabling real-time adaptation and control while guaranteeing
asymptotic tracking convergence. The Lb-KAN achieves improved function
approximation performance and provides visualization of explicit functional
decomposition.

While Lb-KAN provides explicit functional decomposition, this work
clarifies the challenges paving the way for future research to further
achieve symbolic interpretability. Future research could explore introducing
sparsity-inducing regularization terms into the Lyapunov-based control
framework, to achieve convergence and symbolic interpretability for
adaptive control.

\bibliographystyle{ieeetr}
\bibliography{master,ncr,encr,kan_ncr}

\end{document}